\documentclass[structabstract]{aa}
\usepackage{txfonts}
\usepackage{graphicx}
\usepackage{natbib}

\def\beq{\begin{eqnarray}}
\def\eeq{\end{eqnarray}}

\begin{document}

\title{Jet Precession Driven by Neutrino-Cooled Disc for Gamma-Ray Bursts}
\author{Tong Liu\inst{1}\and  En-Wei Liang\inst{2}\and  Wei-Min Gu\inst{3}\and Xiao-Hong Zhao\inst{1} \and  Zi-Gao Dai\inst{1} \and  Ju-Fu
Lu\inst{3}}

\institute{Department of Astronomy, Nanjing University, Nanjing,
Jiangsu 210093, China \and Department of Physics, Guangxi
University, Nanning, Guangxi, 530004, China \and Department of
Physics and Institute of Theoretical Physics and Astrophysics,
Xiamen University, Xiamen, Fujian 361005, China}

\offprints{Tong Liu, \email{ tongliu@nju.edu.cn }}

\titlerunning{Jet Precession Driven By Neutrino-Cooled Disc}
\authorrunning{Liu et al}

\date{Received  2009 / Accepted 2010}

\abstract{}{A model of jet precession driven by a neutrino-cooled
disc around a spinning black hole is present in order to explain the
temporal structure and spectral evolution of gamma-ray bursts
(GRBs).}{The differential rotation of the outer part of a neutrino
dominated accretion disc may result in precession of the inner part
of the disc and the central black hole, hence drives a precessed jet
via neutrino annihilation around the inner part of the disc.}{Both
analytic and numeric results for our model are present. Our
calculations show that  a black hole-accretion disk system with
black hole mass $M \simeq 3.66 M_\odot$, accretion rate $\dot{M}
\simeq 0.54 M_\odot \rm s^{-1}$, spin parameter $a=0.9$ and
viscosity parameter $\alpha=0.01$ may drive a precessed jet with
period $P=1$ s and luminosity $L=10^{51}$ erg  s$^{-1}$,
corresponding to the scenario for long GRBs. A precessed jet with
$P=0.1$s and $L=10^{50}$ erg  s$^{-1}$ may be powered by a system
with $M \simeq 5.59 M_\odot$, $\dot{M} \simeq 0.74 M_\odot \rm
s^{-1}$, $a=0.1$, and $\alpha=0.01$, possibly being responsible for
the short GRBs. Both the temporal and spectral evolution in GRB
pulse may explained with our model.}{GRB central engines likely
power a precessed jet driven by a neutrino-cooled disc. The global
GRB lightcurves thus could be modulated by the jet precession during
the accretion timescale of the GRB central engine. Both the temporal
and spectral evolution in GRB pulse may be due to an viewing effect
due to the jet precession.}

 \maketitle

\section{Introduction}
Internal shock models are extensively discussed for gamma-ray bursts
(GRBs) (Rees \& M\'{e}sz\'{a}ros 1992; M\'{e}sz\'{a}ros \& Rees
1993; Zhang \& M\'{e}sz\'{a}ros 2004), in which an individual shock
episode of two collision shells gives rise to a pulse, and random
superposition of pulses results in the observed complexity of GRB
light curves (e.g., Daigne \& Mochkovitch 1998; Kobayashi et al.
1999). The observed flux rapidly increases in the dynamic timescale
of two shell collision, then decays due to the delayed photons from
high latitudes with respect to the line of sight upon the abrupt
cessation of emission after the crossing timescale, shaping the
observed fast-rise-exponential-decay (FRED) pulses. However, some
well-separated GRB pulses show symmetric structure, and their peak
energy of the $\nu F_\nu$ spectrum ($E_p$) traces the lightcurve
behavior (Liang \& Kargatis 1996; Liang \& Nishimura 2004; Lu \&
Liang 2009; Peng et al. 2009). Both the temporal and spectral
properties of these symmetric pulse are difficult to be explained
with internal shocks. In addition, the observed $E_{\rm iso}-E_p$
relation (Amati et al. 2002) or $L_{\rm iso}-E_p$ relation (Wei \&
Gao 2003; Liang et al. 2004; Yonetoku et al. 2004) also challenge
the internal shock models (e.g., Zhang \& M{\'e}sz{\'a}ros 2002).

Quasi-periodic feature observed in some GRB light curves motivated
ideas that the GRB jet may be precessed (Blackman et al. 1996;
Portegies Zwart et al. 1999; Portegies Zwart \& Totani 2001; Reynoso
et al. 2006; Lei et al. 2007). It is generally believed that the
progenitors of short and long GRBs are the mergers of two compact
objects (Eichler et al. 1989; Paczy\'{n}ski 1991; Narayan et al.
1992; see recent review by Nakar 2007) and core collapsars of
massive stars (Woosley 1993; Paczy\'{n}ski 1998; see reviews by
Woosley \& Bloom 2006), respectively. Although the progenitors of
the two types of GRBs are different, the models for their central
engines are similar, and essentially all can be simply classed as a
rotating black hole with a rapidly hyper-accreting process of a
debris torus surrounding the central black hole. Such a black
hole-disk system drives an ultra-relativistic outflow to produce
both the prompt gamma-rays and afterglows in lower energy bands. The
most popular one is neutrino dominated accretion flows (NDAFs),
involving a black hole of $2\sim10 M_\odot$ and a hyper-critical
rate in the range of $0.01 \sim 10 M_\odot {\rm s}^{-1}$ (Popham et
al. 1999; Narayan et al. 2001; Kohri \& Mineshige 2002; Di Matteo et
al. 2002; Kohri et al. 2005, 2007; Lee et al. 2005; Gu et al. 2006;
Chen \& Beloborodov 2007; Liu et al. 2007, 2008, 2010; Kawanaka \&
Mineshige 2007; Janiuk et al. 2007). The different direction of
angular momentum of two compact objects and the anisotropic
fall-back mass in collapsar may conduct precession between black
hole and disc. In this scenario, the inner part of the disc is
driven by the black hole during the accretion process. The
differential rotation between the inner and outer parts may result
in precession of the inner part of the disc and the central black
hole, hence drive a precessed jet produced by neutrino annihilation
around the inner part of the disc, forming an S- or Z-shaped jet as
observed in many extragalactic radio sources (see, e.g. Florido et
al. 1990).  A tilted accretion disc surrounding a black hole would
also make the precession of the black hole and result in an S-shaped
jet as observed in SS 433 (Sarazin et al. 1980; Lu 1990; Lu \& Zhou
2005), although the angle between angular momentum of black hole and
disc is small due to that evolution of a two compact object system
may decrease the angle between them in mergers or the anisotropic
fall-back mass cannot produce large angle between black hole and
fall-back mass in collapsars.

In this paper, we propose a model of jet precession driven by a
neutrino-cooled disc around a spinning black hole in order to
explain the temporal structure and spectral evolution of GRBs. In
our model, the global profile of a GRB lightcurve may be modulated
by the jet procession. The temporal structure and spectral evolution
may signal an on-axis/off-axis cycle of the light of sight (LOS) to
a precessed jet axis, as proposed by some authors to explain the
nature of low luminosity GRBs 980425 and 031203 (Nakamura 1998;
Eichler \& Levinson 1999; Waxman 2004; Ramirez-Ruiz et al. 2005) or
to present a unified model for GRBs and X-ray flashes (Yamazaki et
al. 2003) and the observed spectral lag in long GRBs (Norris 2002;
Salmonson \& Galama 2002).

We present both analytic and numerical analysis for jet precession
driven by a neutrino-cooled disc around a spinning black hole in
sections 2 and 3. Simplifying the jet emission surface as a point
source, we demonstrate the profile and evolution of a GRB pulse in
section 4. Conclusions and discussion are shown in section 5.

\section{Model}

\begin{figure}
\centering
\includegraphics[width=0.4\textwidth]{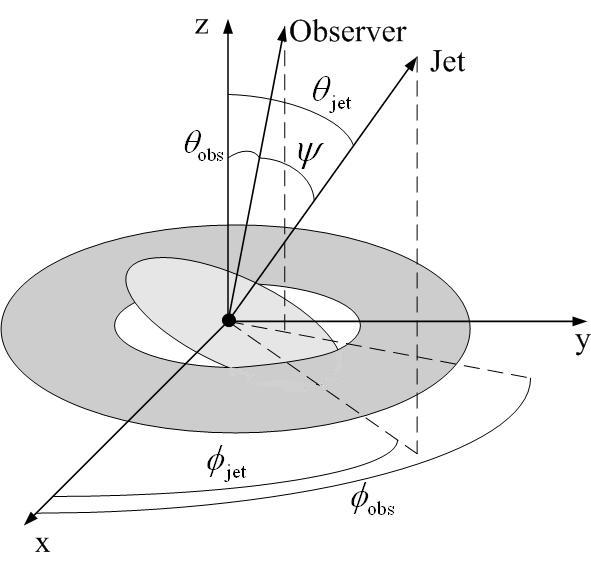}
\caption{Schematic picture of a precessing system.}
\label{sample-figure}
\centering
\end{figure}

An accretion disk would be warped by its precession (Sarazin et al.
1980). We consider a spinning black hole surrounding a tilted
accretion disc that its rotation axis is misaligned with that of the
black hole, as shown in Fig. 1. Its angular momentum is $J_\ast = aG
M^2 /c$, where $M$ is the black hole mass and $a$ ($0<a<1$) is the
dimensionless specific angular momentum. Since $dJ = 2 \pi r^2
\Sigma v_\phi dr$ for a ring at radius $r$ in the disc with width
$dr$, we get $J(r) = dJ/d(\ln r) = 2 \pi r^3 \Sigma v_\phi$, where
$\Sigma$ and $v_\phi$ are the disk surface density and rotational
velocity. Due to the Lense-Thirring effect (Lense \& Thirring 1918),
the disc material interior to a critical radius $r_p$, which is
defined as $J(r_p) = J_\ast$, will be aligned with the equatorial
plane of the black hole. The outer portion of the disc ($r \ga r_p$)
with sufficiently large angular momentum keeps its orientation. This
makes the black hole precess along with the inner disc (Bardeen \&
Petterson 1975). A jet dominated by the ejections of neutrino
annihilation around the inner part of the disc thus would be
precessed (Popham et al. 1999; Liu et al. 2007). The precession rate
$\Omega$ of the central black hole and the inner disc is given by
$\Omega=2GJ(r)/c^2 r^3 $ (Sarazin et al. 1980). Note that regions
with $r \ga r_p$ in the disc should contribute to the precession.
The $\Omega$ decreases as $r$ increases, so one cannot expect a
period behavior in an observed light curve from our model.

With the continuity equation \beq \dot{M}=-2 \pi r \Sigma v_r, \eeq
the precession period $P$ then can be expressed as \beq\ P \equiv
\frac{2 \pi} {\Omega} = (\pi M)\Big(\frac{a}{G}\Big )^{\frac{1}{2}}
\Big(\frac{c v_r }{\dot{M}v_\phi} \Big )^{\frac{3}{2}}, \eeq where
$\dot{M}$ is the accretion rate and $v_r$ the radial inflow velocity
of the disc material. Assuming that the angular velocity is
approximately Keplerian, we have $v_\phi = r \Omega_{\rm K}$, and
the vertical scale height of the flow can be written as $H=c_s /
\Omega_{\rm K}$, where $\Omega_{\rm K}= (GM/r^3)^{1/2}$ and  $c_s$
are the Keplerian angular velocity and the sound speed,
respectively. The $v_r$ can be estimated as $v_r \sim \alpha c_s (H
/ r)$ (Kato et al. 2008), where $\alpha$ is the constant viscosity
parameter of the disc. Substituting the expressions of $v_\phi$ and
$v_r$ into equation (3), we have \beq\ P = 1.42 \times 10^{3}
a^{\frac{1}{2}} \alpha^{\frac{3}{2}}\Big(\frac{M}{M_\odot}\Big)
\Big({\frac{\dot{M}}{M_\odot \rm s^{-1}}}\Big)^{-\frac{3}{2}}
\Big(\frac{H}{r}\Big)^3~ \rm s. \eeq\ It is found that $P$ is
sensitive to $\alpha$, $\dot{M}$ and $H/r$. These parameters are
time-dependent in the GRB phenomenon; hence the precession period
should evolve with time. Regions with $r \ga r_p$ in the disc should
contributes to the precession, and the evolution of the
hyper-accretion process would make the mass and the angular momentum
of black hole increase (Belczy\'nski et al. 2008; Janiuk et al.
2008), hence make evolution of the precession period. In addition,
the nutation in the accretion system even makes the observed profile
be much complicated. Therefore, one cannot expect clear period
information in the GRB lightcurves. If the periods are shorter than
the accretion timescale, the observed lightcurve may be composed of
some pulses. Occasionally, the lightcurves may show quasi-periodic
feature, such as that observed in BATSE trigger 1425 (Portegies
Zwart et al. 1999). If the periods are longer than the accretion
timescale, the global lightcurve may be a broad pulse. This is
different from that for SS433 (Sarazin et al. 1983), in which it is
assumed that the precession periods are shorter than the viscous
timescale without rapidly evolving with time, hence the periodic
lightcurve is a natural consequence in SS 433.

\begin{figure}
\centering
\includegraphics{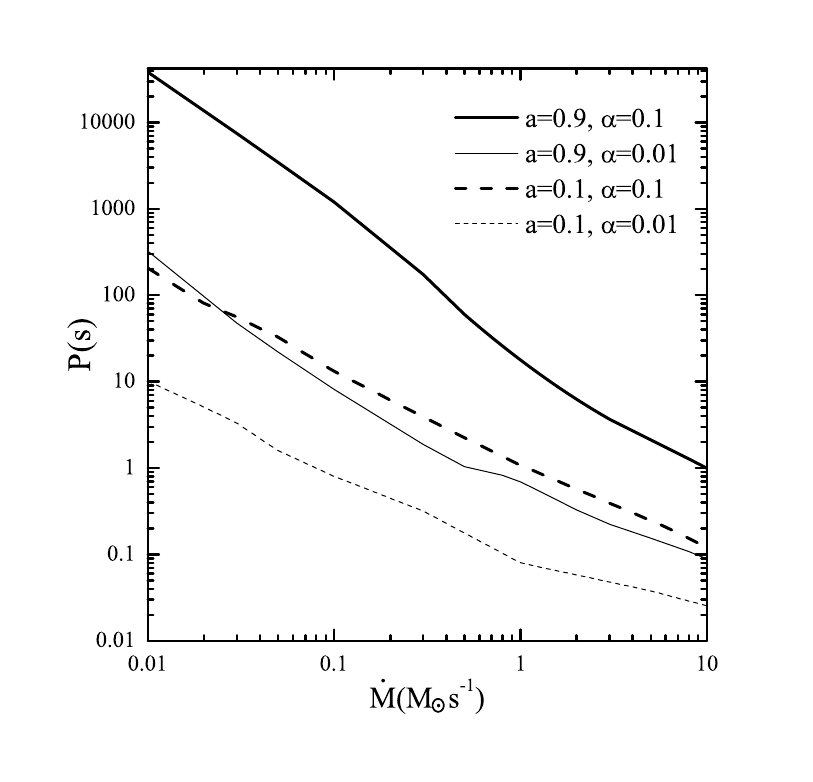}
\caption{Illustration of numerical results for $P$ as a function of
$\dot{M}$ using different parameter sets as marked in the plot.}
\label{sample-figure}
\centering
\end{figure}

\section{Numerical Results}
The Eq. (3) show an explicit dependence of $P$ to $a$, $\alpha$,
$\dot{M}$, $M$, and $H/r$. However, the thickness of NDAF also
depends on $M$, $\dot{M}$, $a$ and $\alpha$. Similar to $P$, the
injected neutrino annihilation luminosity $L$ is also a function of
these parameters. In order to illustrate both the dependences of $P$
and $L$ on these parameters, we present numerical calculation with
the method by Riffert \& Herold (1995). This method defines general
relativistic correction factors to simulate the precession period
related to the spin of a black hole. They are written as \beq\ A =
1-\frac{2GM}{c^2 r}+\Big(\frac{a G M}{c^2 r}\Big)^2, \eeq\ \beq\ B =
1- \frac{3GM}{c^2 r}+2a \Big(\frac{a G M}{c^2 r}\Big)^{\frac{3}{2}},
\eeq\ \beq\ C = 1- 4a \Big(\frac{a G M}{c^2
r}\Big)^{\frac{3}{2}}+3\Big(\frac{a G M}{c^2 r}\Big)^2 \eeq\ \beq\ D
= \int_{r_{ms}}^{r} \frac{\frac{x^2 c^4}{2G^2}-\frac{3xMc^2}{G}+
4(\frac{x a^2 M^3 c^2}{G})^{\frac{1}{2}}-\frac{3a^2
M^2}{2}}{(xr)^{\frac{1}{2}}[\frac{x^2 c^4}{G^2}- \frac{3xMc^2}{G}+2
(\frac{x a^2 M^3 c^2}{G})^{\frac{1}{2}}]} dx,\eeq where $r_{ms}$ is
the inner boundary of the disc. The equation of conservation of mass
remains valid, while hydrostatic equilibrium in the vertical
direction leads to a corrected expression for the half thickness of
the disc (Riffert \& Herold 1995), \beq H \simeq c_{\rm
s}\Big(\frac{ r^3}{G M}\Big)^{\frac{1}{2}}
\Big(\frac{B}{C}\Big)^{\frac{1}{2}}, \eeq where $c_{\rm
s}=(p/\rho)^{1/2}$, $p$ and $\rho$ are the total pressure and
density of the disk, respectively. The viscous shear $T_{r \phi}$ is
also corrected as \beq\ T_{r \phi} = - \alpha p
\Big(\frac{A}{BC}\Big)^{\frac{1}{2}}, \eeq and the angular momentum
equation can be simplified as (Riffert \& Herold 1995, Lei et al.
2009) \beq\ T_{r \phi} = \frac{\dot{M}}{4\pi H
}\Big(\frac{GM}{r^3}\Big)^{\frac{1}{2}}
\Big(\frac{D}{A}\Big)^{\frac{1}{2}}.\eeq The equation of state is
\beq\ p = p_{\rm gas} + p_{\rm rad} + p_{\rm e} + p_\nu, \eeq where
$p_{\rm gas}$, $p_{\rm rad}$, $p_{\rm e}$, and $p_\nu$ are the gas
pressure from nucleons, radiation pressure of photons, degeneracy
pressure of electrons, and radiation pressure of neutrinos,
respectively (see, e.g. Di Matteo et al. 2002; Liu et al. 2007). The
energy equation is written as \beq\ Q_{\rm vis} = Q_{\rm adv} +
Q_{\rm photo} + Q_\nu, \eeq where $Q_{\rm vis}$, $Q_{\rm adv}$,
$Q_{\rm photo}$ and $Q_\nu$ are the viscous heating rate, the
advective cooling rate, the cooling rate due to photodisintegration
of $\alpha$-particles and the cooling due to the neutrino radiation,
respectively (see, e.g. Di Matteo et al. 2002; Liu et al. 2007). The
heating rate $Q_{\rm vis}$ is expressed as \beq\ Q_{\rm vis} =
\frac{3GM \dot{M}}{8 \pi r^3} \frac{D}{B}. \eeq\

\begin{figure}
\centering
\includegraphics{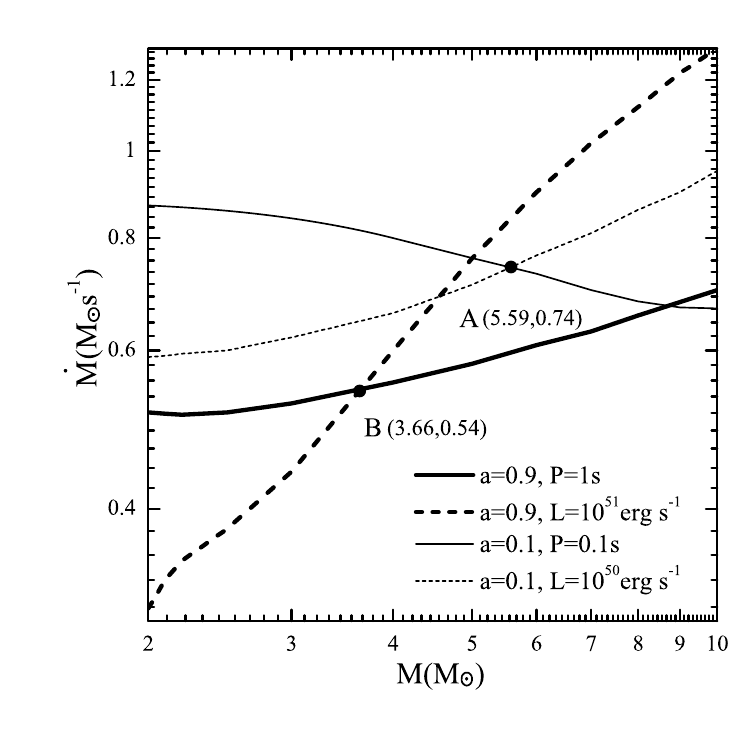}
\caption{$\dot{M}$ as a function of $M$ for different parameter sets
as marked in the plot. The viscosity parameter is adopted as $\alpha
= 0.01$. The Two interaction points A nd B indicate estimates of $M$
and $\dot{M}$ for a given set of ($P$, $L$, $a$, $\alpha$).}
\label{sample-figure}
\centering
\end{figure}

The equation system consisting of Eqs. (1), (2), (4)-(13) is closed
for an unknown precession period $P$. It can be numerically solved
for a given parameter set of $M$, $\dot{M}$, $a$, and $\alpha$. We
show $P$ as a function of $\dot{M}$ for the parameter sets ($a=0.9$,
$M = 3 M_\odot$, $\alpha=0.01$), ($a=0.9$, $M = 3 M_\odot$,
$\alpha=0.1$), ($a=0.1$, $M = 3 M_\odot$, $\alpha=0.1$) and
($a=0.1$, $M = 3 M_\odot$, $\alpha=0.01$) in Fig. 2. It is found
that $P$ varies from tens of milliseconds to 10 ks, if
$\dot{M}=0.01\sim 10$ $M_\odot$/s, $\alpha=0.01\sim 0.1$, and
$a=0.1\sim 0.9$. It can approach the timescale of lightcurve or be
longer than the accretion timescale whose provide a couple of
possibilities of lightcurve. In collapsar scenario, the central
black hole would be rapidly rotates, i.e., $a\gtrsim 0.9$. For the
compact object mergers, the spin of the black hole is not strictly
as high as that in the collapsar scenario (e.g., van Putten et al.
2001). Assuming that the global GRB lightcurves are modulated by the
jet precession during the accretion timescale of the GRB central
engine, the profile of a pulse duration may be comparable to $P$.
Statistical analysis shows that the typical durations of long and
short GRB pulses are $\sim 1$ and $0.1$ second, respectively (Liang
et al. 2002; Nakar \& Piran 2002). From Fig. 2, we find that the
case of ($\alpha=0.1$, $a=0.9$) yields a $P$ value much larger that
$1$ seconds in the range of $\dot{M}=0.01\sim 10 M_\odot$. For the
case of ($\alpha=0.01$, $a=0.9$), we get $P=0.1\sim 1$ seconds for
$\dot{M}=0.4\sim 10 M_\odot$. This is consistent with the observed
pulse durations for long GRBs. In order to explain the duration of
short GRB pulses, our model requires lower $a$ and $\alpha$ as well
as higher $\dot{M}$ than that for the long GRBs, indicating that the
short duration GRBs may be powered by much violent accretion process
than the long ones.

\begin{figure*}
\centering
\includegraphics[angle=0,scale=0.70]{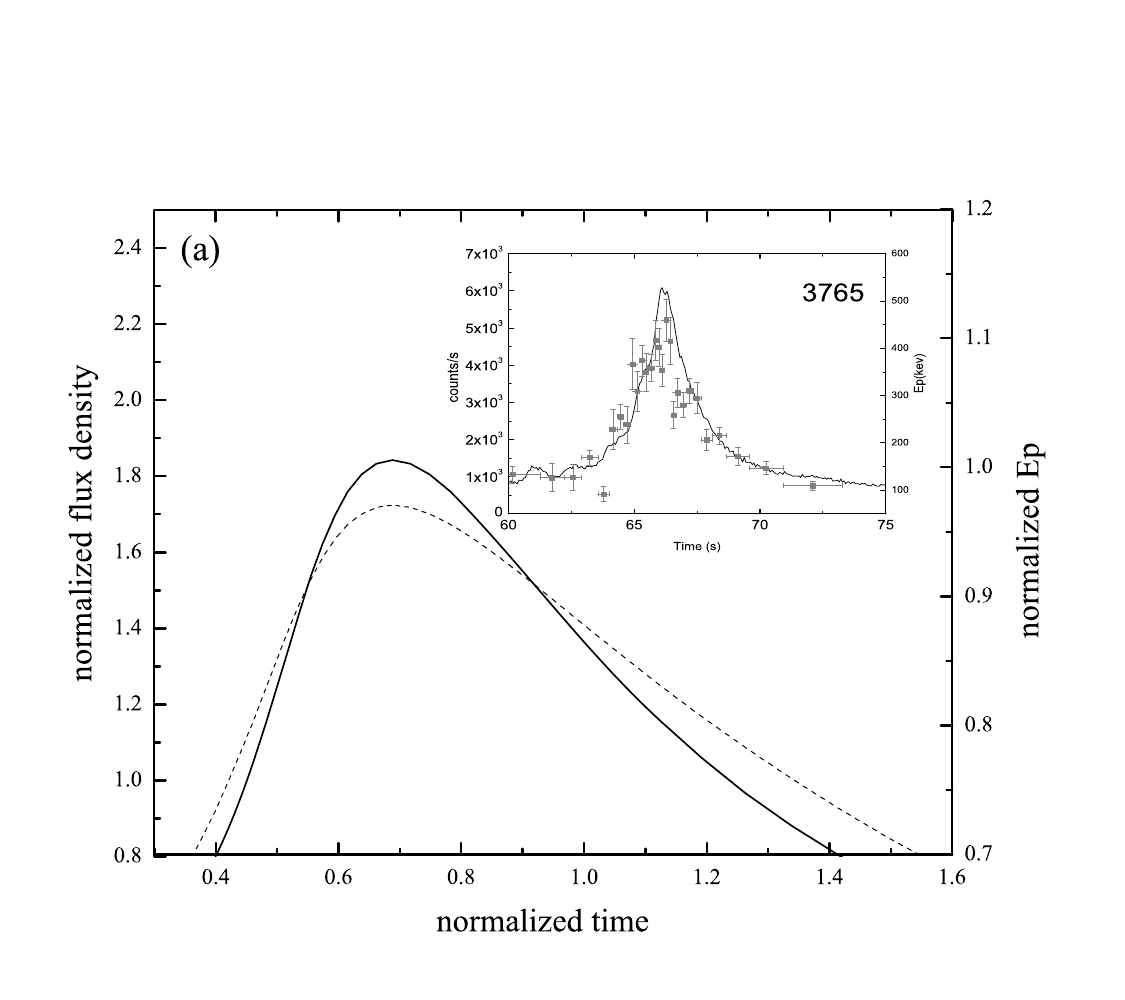}
\includegraphics[angle=0,scale=0.75]{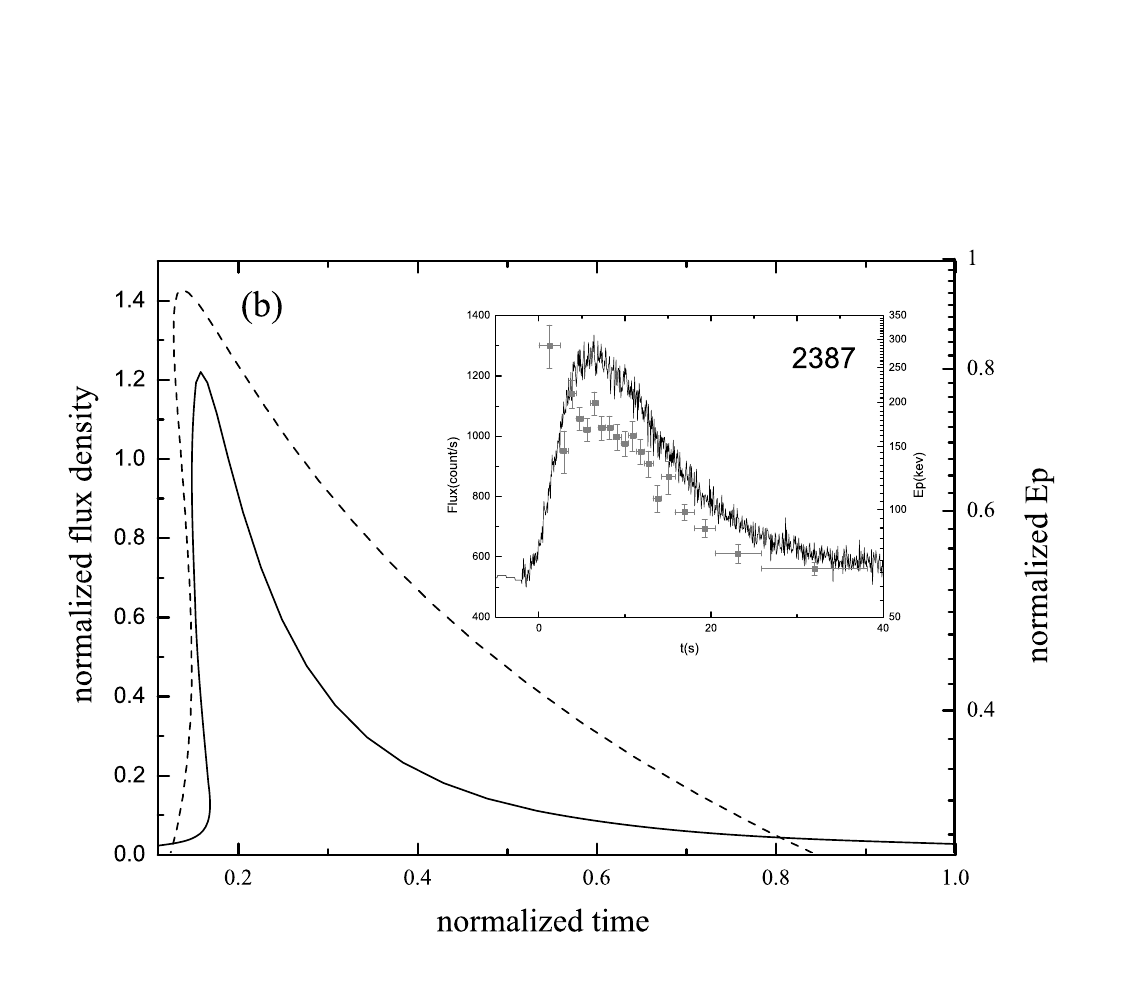}
\caption{Predicted flux $F$ (the solid line) and $E_p$(the dashed
line) with our model for a symmetric pulse (panel a)and a FRED pulse
(panel b) with comparisons to the observations(insets).}
\label{sample-figure}
\centering
\end{figure*}

The observed luminosity of prompt gamma-rays may also place
constraints on our model parameters. We assume that the observed
gamma-ray luminosity is comparable to the injected neutrino
annihilation luminosity $L$. Similar to $P=P(M, \dot{M}, a,
\alpha)$, $L$ is a function of $M$, $\dot{M}$, $a$, and $\alpha$,
written as $L=L(M, \dot{M}, a, \alpha)$. It can be calculated
following the approach of Ruffert et al (1997), Popham et al (1999),
Rosswog et al (2003), and Liu et al (2007). Since the calculation of
$L$ (or $P$) as a function of these parameters is a huge
time-consumed task, we perform our calculation for typical $L$
values only and present our results with $\dot{M}$ as a function of
$M$, $a$, $\alpha$ for a given $L$ (or $P$). We take $L=10^{51}$ erg
s$^{-1}$ for long GRBs and $L=10^{50}$ erg s$^{-1}$ for short GRBs.
Based on our analysis above, we also calculate $\dot{M}$ as a
function of $M$ for $P=1$ s and $P=0.1$ s in case of parameter set
($L$, $a$, $\alpha$)=($10^{51}$ erg s$^{-1}$, 0.9, 0.01) and ($L$,
$a$, $\alpha$)=($10^{50}$ erg s$^{-1}$, 0.1, 0.01). We show
$\dot{M}$ as a function of $M$ for different parameter sets in Fig.
3. It is found that for a given luminosity, $\dot{M}$ as a function
of $M$ greatly depends on the rotation of the black hole (see the
dotted lines in Fig. 3). The accretion rate $\dot{M}$ does not
significantly increase with $M$ for $a=0.1$. However, it rapidly
increases with $M$ for $a=0.9$. The behavior of the function
$\dot{M}(M)$ varies for different $P$ (see the solid lines in Fig.
3). As shown in Fig. 3, $\dot{M}$ slightly increase with $M$ if
$P=1$ second. However, it even decreases with $M$ for $P=0.1$
second. This conflicts with that shown in the explicit form of $P$
(see Eq. 3). The reason is that the thickness of NDAF depends on
$M$, $\dot{M}$, $a$ and $\alpha$ (see Fig. 9 in Liu et al. 2007).
One cannot expect an explicit dependence between $\dot{M}$ and $M$
for a given $P$. The intersections between the lines are estimates
of $M$ and $\dot{M}$ for a given set of ($P$, $L$, $a$, $\alpha$).
We find that $M \simeq 3.66 M_\odot$ and $\dot{M} \simeq 0.54
M_\odot \rm s^{-1}$ for $P=1$ s, $L=10^{51}$ erg s$^{-1}$, and
$a=0.9$, corresponding to the scenario for long GRBs. For $P=0.1$s,
$L=10^{50}$ erg s$^{-1}$, and $a=0.1$, we get $M \simeq 5.59
M_\odot$ and $\dot{M} \simeq 0.74 M_\odot \rm s^{-1}$, corresponding
to the scenario for short GRBs. These solutions are generally
consistent with the requirements of GRB productions in simulations
for collapsars (e.g. MacFadyen \& Woosley 1999) and for binary
coalescence of a neutron star and a black hole or two neutron stars
(e.g. Klu\'{z}niak \& Lee 1998).

\section{Temporal Profile and Spectral Evolution of a GRB Pulse from a Precessing Jet}
As discussed above, in the framework of our model one cannot expect
period information from the observed lightcurves since the
precession period is time-dependent. Since the period is a function
of some time-dependent parameters as mentioned above, temporal
profile and spectral evolution of pulses in GRB lightcurves may be a
direct information of jet precession since the jet precession may
conduct an on-axis/off-axis cycle during a precession period for a
given observer.

As discussed in Section 1, the $E_p$-tracing-flux spectral evolution
feature is observed in some GRB pulses (e.g. Liang et al. 1996; Peng
et al. 2009; Lu \& Liang 2009). The profiles of these pulses are
generally FRED, and occasionally are symmetrical. These temporal and
spectral features can be explained with our model. We just
illustrate the lightcurve and the spectral evolution for a point
source with arbitrary radiation intensity in the axis with an
arbitrary precession period with ultra-relativistic velocity in the
jet axis. As shown by Granot et al. (2002), assuming the emitting
region as a point source in the jet axis, the calculation can give
reasonable results without any assumption on the jet structure.
Therefore, we adopt the point source assumption in our calculations.
We just illustrate the lightcurve and the spectral evolution for a
point source with arbitrary radiation intensity in the axis with an
arbitrary precession period for an observer (on-axis and off-axis)
at the rest frame in Section 4. If the emitting region is a shell of
the jet with certain opening angle, the peak of the pulse would be
flattened in case of a uniformed jet. Our calculation is followed by
that present in Granot et al. 2002.

The observed flux $F$ and $E_p$ would be amplified due to the
Doppler effect, $F= F_0(1-\beta)^3/(1-\beta \rm cos \Psi) ^3$, $E_p=
E_{p0}(1-\beta)/(1-\beta \rm cos \Psi) $. The observed time scale
would be $t= t_0 (1-\beta \rm cos \Psi)/(1-\beta)$, where the
subscript 0 means the ``on-axis'' quantities, $\Psi$ is the view
angle between the jet axis and the LOS and $\beta=
(\Gamma^2-1)^{1/2}/\Gamma$, $\Gamma$ is the Lorentz factor. From
Fig. 1, we have \beq {\rm cos} \Psi = {\rm cos} \theta_{\rm jet}
{\rm cos} \theta_{\rm obs} + {\rm sin} \theta_{\rm jet} {\rm sin}
\theta_{\rm obs}{\rm cos} (\phi_{\rm jet} - \phi_{\rm obs}),\eeq
where $\phi_{\rm jet} = \phi_{\rm jet, 0} +2 \pi t_1/P$ ($t_1$ is
the time in the rest frame of the central engine) and $z$-axis in
Fig. 1 is the direction of angular momentum of the outer part. We
assume $\Gamma=300$ in our calculations. We compute the observed
flux and peak energy of the $\nu F_\nu$ spectrum in a precession
period corresponding to an observed pulse. Figure 4 demonstrates the
initial ``off-axis" and initial ``on-axis" lightcurves and
corresponding $E_p$ evolutions for an point source with arbitrary
flux intensity and spectral hardness in the jet with arbitrary
precession period $P$. The adopted parameters are $\theta_{\rm
obs}=1^\circ$, $\phi_{\rm obs} = 90^\circ$, $\theta_{\rm jet} =
4^\circ$, and $\phi_{\rm jet, 0} = 0^\circ$ for initial ''off-axis''
observer and $\theta_{\rm obs}= 1^\circ$, $\phi_{\rm obs} =
90^\circ$, $\theta_{\rm jet} = 2^\circ$ and $\phi_{\rm jet, 0} =
0^\circ$ for initial ``on-axis" observers. Two samples of the
observations are also shown in Fig. 4 for comparisons. It is found
that our model can re-produce both FRED and symmetric pulses with
$E_p$-tracing-flux behavior, depending on the initial on-axis of
off-axis observations.

\section{Conclusions}
We have suggested that the differential rotation of the outer part
of a neutrino dominated accretion disc may result in precession of
the central black hole and the inner part of the disc, hence may
power a precessed jet via neutrino annihilation around the inner
part of the disc. Both analytic and numeric results are present. Our
calculations show that for a  black hole-accretion disk system with
$M \simeq 3.66 M_\odot$, $\dot{M} \simeq 0.54 M_\odot \rm s^{-1}$,
$a=0.9$ and $\alpha=0.01$ may drives a precessed jet with $P=1$ s
and $L=10^{51}$ erg  s$^{-1}$, corresponding to the scenario for
long GRBs. A precessed jet with $P=0.1$s and $L=10^{50}$ erg
s$^{-1}$ may be powered by a system with  $M \simeq 5.59 M_\odot$,
$\dot{M} \simeq 0.74 M_\odot \rm s^{-1}$, $a=0.1$, and
$\alpha=0.01$, possibly being responsible for the short GRBs. These
results are generally consistent with simulations for long and short
GRB productions from collapsars and from mergers of compact stars.
Both temporal and spectral features observed in GRB pulses may be
explained with our model.

The correlation between $E_{\rm iso}$ (or $L_{\rm iso}$) and $E_p$
in the burst frame (Amati et al. 2002; Liang et al. 2004) are
difficult to be explained in the framework of internal shock
scenarios. Our model suggests an $E_p$-tracing-flux behavior within
a GRB pulse due to the on-axis/off-axis effect for a given observer,
similar to that proposed by Yamazaki et al. (2004). The
$E_p$-tracing-flux behavior would give rise to the observed
correlations between $E_{\rm iso}$ ($L_{\rm iso}$) and $E_p$ in the
burst frame.

\section*{acknowledgments}
We thank the anonymous referee for very useful comments. We also
thank Bing Zhang, Shuang-Nan Zhang, Li-Xin Li, and Wei-Hua Lei for
beneficial discussion. This work was supported by the China
Postdoctoral Science Foundation funded project 20080441038 (T.L.),
the National Natural Science Foundation of China under grants
10778711 (W.M.G.), 10833002 (J.F.L. and W.M.G.), 10873002 (E.W.L.),
10873009 (Z.G.D.), the National Basic Research Program (973 Program)
of China under Grant 2009CB824800 (E.W.L., W.M.G., and J.F.L.). E.
W. L. also acknowledges the support from Guangxi SHI-BAI-QIAN
project (Grant 2007201), the program for 100 Young and Middle-aged
Disciplinary Leaders in Guangxi Higher Education Institutions, and
the research foundation of Guangxi University(M30520).

\clearpage

\end{document}